\documentclass[aps,prd,showpacs,twocolumn,nofootinbib,superscriptaddress]{revtex4-2}
\usepackage{amsmath,amssymb,amsfonts}
\usepackage{bm}
\usepackage{verbatim}
\usepackage{hyperref}
\usepackage{slashed,braket}
\usepackage{color}
\usepackage{graphicx,graphics}
\usepackage[title,titletoc]{appendix}
\usepackage{stackengine}
\usepackage{mathrsfs}

\newcommand{\cA}{\mathcal{A}}
\newcommand{\cD}{\mathcal{D}}

\newcommand{\sD}{\slashed{D}}

\newcommand{\tr}{\operatorname{tr}}

\newcommand{\Pexp}{\operatorname{Pexp}}

\newcommand{\w}{\omega}

\begin{document}

\title{Non-perturbative suppression of the Chiral Magnetic Effect in Quark Gluon Plasma}

\author{Ruslan A.~Abramchuk}
\email{abramchukrusl@gmail.com}
\email{ruslanab@ariel.ac.il}
\affiliation{Physics Department, Ariel University, Ariel 40700, Israel}

\date{\today}

\begin{abstract}
    
    In this paper the non-perturbative suppression of the Chiral Magnetic Effect (CME) is investigated in the deconfined region of the QCD phase diagram (as a model for Quark-Gluon Plasma (QGP) emerging in Heavy Ion Collisions (HIC)),
    using the Kubo formula to calculate a linear response to the chiral imbalance,
    and the Field Correlator Method to address the strong interaction in QCD. 

The estimate suggests that CME is severely suppressed in most of the QCD phase diagram, 
    except within a narrow strip at baryon densities above nuclear 
    and temperatures just above the deconfinement transition. 
The result suggests refining the conditions under which the CME might be observable ---
    prioritizing the conditions at lower energy RHIC-BES, SPS and upcoming FAIR, NICA, J-PARC-HI,
    which supposedly produce high baryon density QGP,
    over the higher energy RHIC and LHC,
    which produce high temperature QGP. 
\end{abstract}

\maketitle

\section{Introduction}
\label{SectIntro}

The prominent Chiral Magnetic Effect \cite{Fukushima2008}, 
    which suggest emergence of electric current along magnetic field in response to a chiral imbalance, 
    is a promising probe for QGP and magnetic field formed in high energy HIC. 
However, experimental detection of CME remains elusive.
A comprehensive set of theoretical approaches to CME in interacting system
    (including Lattice QCD, AdS-CFT correspondance, anomalous hydrodynamics, phenomenological analysis)
    which is notable and deemed to be successful \cite{Kharzeev2024CMEHIC},
points toward further analysis of HIC data from RHIC and LHC at higher and higher collision energies,
    where stronger magnetic fields and more ample chiral imbalance fluctuations might emerge,
    while CME itself is regarded as immutable due to its assumed connection to the chiral anomaly.

\begin{figure}
    \center{\includegraphics[width=\linewidth]{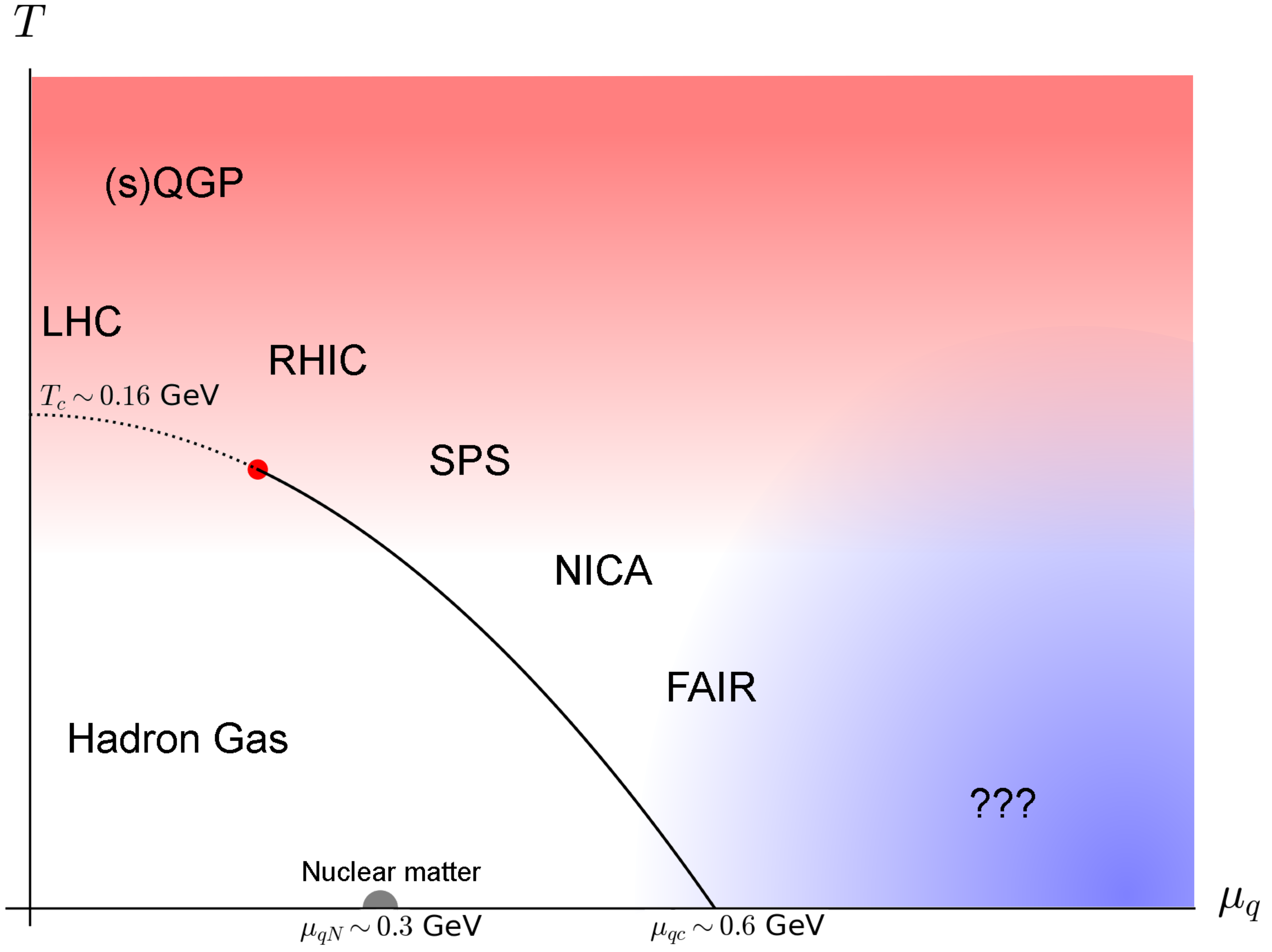}}
    \caption{
        QCD phase diagram sketch. 
        Quark chemical potential \(\mu_q\) is related to the baryon chemical potential \(\mu_B\) as \(\mu_B=3\mu_q,~\mu_q=\mu_u=\mu_d,~\mu_s=0\).
    \label{FigPhaseDiagram}
    }
\end{figure}

Heavy-ion collision experiments span a range of baryon density and temperature conditions, 
    providing diverse probes of QCD matter. 
The conditions achieved in existing and upcoming experiments are mapped onto the phase diagram in Fig.\ref{FigPhaseDiagram}.
At RHIC, the Beam Energy Scan (BES) program explored 
    center-of-mass per nucleon collision energies from \(\sqrt{s_{NN}} = 7.7\) to 200 GeV, 
    achieving baryon chemical potentials \(\mu_B\) up to \(\sim 400\) MeV 
    and temperatures around \(150-170\) MeV at lower energies (e.g., 7.7-19.6 GeV), corresponding to baryon densities of approximately 5-10 times normal nuclear density 
    \(n_0 \approx 0.16~\text{fm}^{-2}\) (which corresponds to baryon chemical potential \(\mu_{BN}\sim 0.9\) GeV and quark chemical potential \(\mu_{qN}\sim 0.3\) GeV), 
    while higher energies (e.g., 200 GeV) produce hot QGP with temperatures exceeding 300 MeV and near-zero net baryon density at mid-rapidity \cite{STAR2017,Andronic2017}. 
The LHC, operating at \(\sqrt{s_{NN}} = 2.76-5.02\) TeV, generates QGP with 
    temperatures exceeding 400 MeV and negligible baryon density (\(\mu_B \sim 0-20\) MeV) at mid-rapidity, 
    though fragmentation regions might exhibit higher baryon densities  \cite{ALICE2010,CMS2013}.

Upcoming experiments, such as the Compressed Baryonic Matter (CBM) experiment at the Facility for Antiproton and Ion Research (FAIR), 
    target \(\sqrt{s_{NN}} = 2.7-4.9\) GeV, 
    probing extremely high baryon densities (up to \(10-12 ~n_0\)) or with \(\mu_B\) up to 600-800 MeV
    and moderate temperatures (\(T \sim 100-150\) MeV), 
    conditions relevant to neutron star interiors \cite{CBM2016}. 
Similarly, the J-PARC Heavy-Ion (J-PARC-HI) project plans to operate at \(\sqrt{s_{NN}} = 5-10\) GeV, 
    expecting baryon densities up to \(5-7 ~n_0\) (\(\mu_B \sim 300-500\) MeV)
    and temperatures of 120-160 MeV, 
    offering insights into the high-density QCD phase diagram \cite{Sako2021}. 

In this work, we estimate the non-perturbative suppression of the CME in hot deconfined QCD.
To avoid questions regarding the nature of chiral imbalance 
    (whether the notion of chiral chemical potential as equilibrium characteristic is well-defined or not)
we employ the Kubo formula, 
    which allows to calculate a linear response to a chiral imbalance over the well-defined thermodynamic equilibrium QCD,
    as suggested in \cite{Brandt2025CMEKubo}.
To address the problem of strong interaction in deconfined QCD we employ the Field Correlator Method 
    (FCM, see \cite{Simonov2018cbk,Lukashov2025} for a recent review). 
By integrating these methods, we quantify the CME’s suppression due to non-perturbative interactions in hot deconfined QCD, 
    building on prior work on the Chiral Vortical and Separational Effects suppression \cite{Zubkov2023,Abramchuk2023CVE,Abramchuk2024CE,Khaidukov2023CSE}. 
However, we do not discuss the characteristic magnitudes of the magnetic field and chiral imbalance 
    as functions of collision energy.

This paper aims to enhance theoretical understanding of CME 
    in hot deconfined QCD (as a model for QGP emerging in HIC)
    phase diagram (in baryon chemical potential -- temperature plane), 
    offering insights into its detectability across HIC energies.

The paper is organized as follows. 
In the next Section we outline the theoretical framework, 
    combining the Kubo formula and FCM to assess the CME suppression in hot dense QCD. 
In Conclusion, we discuss the numerical results for the CME conductivity and its relevance to HIC experiments. 
In the Appendix the relevant approaches of Field Correlator Method 
    are discussed and the relevant literature provided, 
    the FCM inputs are listed. 

\section{CME in the World-line representation}

We apply the Kubo formula for light quarks in QCD to calculate the linear response to the chiral chemical potential,
    and retain only the linear contribution in powers of magnetic field.
The leading non-perturbative effects in deconfined QCD in thermodynamic equilibrium with non-zero temperature \(T=\beta^{-1}\) and quark chemical potential \(\mu\) are treated with FCM.

For a fermion with electric charge \(q_f\) CME in the static limit \cite{Brandt2025CMEKubo}
\begin{gather}
    \vec j_f = \lim_{\w\to+0} q_f^2\vec H\mu_5(\w)I_\text{CME}(\w).
\end{gather}
We calculate \(I_\text{CME}=\lim_{\w\to+0}I_\text{CME}(\w)\) 
    using the world-line (Feynman-Fock-Schwinger) representation as customary in the Field Correlator Method
\begin{gather}
    qHI_\text{CME} = \lim_{\w\to 0}\Pi_{z5}(\omega) = \int d^4x\braket{j_z(0)\rho_5(x)}^\text{reg} \\
    = \int d^4x\braket{\tr_{c,D}\gamma_3S_{0,x}\gamma_4\gamma_5S_{x,0}}^\text{reg}_B \\
    = \int d^4x\langle\tr_{c,D}(\sD_{z=0}+im)(\sD_{\bar z=x}+im) \\
    \iint_0^{+\infty}ds d\bar s ~\zeta(s)\zeta(\bar s)
        (\overline{\cD^4z})^s_{0x}~(\overline{\cD^4\bar z})^s_{x0}\nonumber\\
        e^{-m^2(s+\bar s)-K-\bar K} 
    \gamma_3\Phi_\sigma[z]\gamma_4\gamma_5\Phi_\sigma[\bar z]\rangle_B \nonumber
\end{gather}

With the backgrond gluonic field and the external electromagnetic fields, 
    the covariant derivatives ans the parallel transporters read
\begin{gather}
    \sD = \gamma_\nu(\partial_\nu -igB_\nu^a t_a -iq\cA_\nu),
        \quad \cA_\nu = (0,Hx_1,0,-i\mu/q) \\
    \Phi_\sigma[z] = \Pexp\Big(ig\oint B\cdot dz + iq\oint \vec\cA\cdot d\vec z
        +\mu\int dz_4 + \nonumber\\ 
    + \int_0^s d\tau g\sigma_{\mu\nu}F^{\mu\nu}(z,z_0|B) + 2qHs\sigma_{12}\Big) 
\end{gather}
we expand the spin part of the parralel transporter.
The \(\gamma_5\) matrix makes most of the terms in the Dirac trace vanish since
\begin{gather} 
    \tr_D\gamma_5\gamma_\mu\gamma_\nu\gamma_\lambda\gamma_\rho = -4\epsilon_{\mu\nu\lambda\rho},
        ~\quad \epsilon_{1234}=1.
\end{gather}
The color-magnetic contribution vanishes 
    since the color-magnetic field strength has no preferred direction
    (in contrast to the external magnetic field).
The covariant derivatives vanish {because of the parallel transporter definition} \cite{Simonov2002vva,Feynman1951}
and the derivatives that act on the path integrals endpoints vanish because of the symmetry in the transverse plane (similar to the consideration of CSE \cite{Zubkov2023}). 

With the defining path integral property
\begin{gather}
    \int d^4x (\overline{\cD^4z})^s_{0x}~(\overline{\cD^4\bar z})^s_{x0}
        = (\overline{\cD^4z})^{s+\bar s}_{00}
\end{gather}
and 
\begin{align}
    \iint_0^{+\infty}&ds d\bar s ~f(s+\bar s) = \\
    =&\int_{-\infty}^{+\infty}d\tau\iint_0^{+\infty}ds d\bar s
        ~\delta(\tau-s-\bar s)f(\tau) \nonumber\\
    =& \int_{-\infty}^{+\infty}d\tau\int_0^{+\infty}ds ~f(\tau)\Theta(\tau-s)
        = \int_0^{+\infty}s ds~ f(s)\nonumber
\end{align}
we obtain
\begin{gather}
    I_\text{CME} = -m^2 
        \int_0^{+\infty}sds~\zeta(s)(\overline{\cD^4z})^s_{00} e^{-m^2s-K} \times \label{EqContLoop}\\
    \times\langle\tr_{c,D}\Phi[z]\gamma_3\gamma_4\gamma_5 ~2s\sigma_{12}\rangle_B \nonumber \\
    = {8}m^2\int_0^{+\infty}s^2ds~\zeta(s)(\overline{\cD^4z})^s_{00} e^{-m^2s-K}
        \langle\tr_{c}\Phi[z]\rangle_B \label{EqPathInt}
\end{gather}

The path integrals are calculated following \cite{Zubkov2023,Abramchuk2023CVE}.
The non-perturbative effects are parametrized by the singlet free energy \(V_1(T)\) 
    and the screened quark mass \(\bar M(T)\)  listed in the Appendix \ref{SectFCM}.
The analytical form is a rough approximation
    that, on the other hand, allows us to proceed and obtain an analytic final result.
With all the approximations involved,
    especially \(V_1(T,\mu)\to V_1(T)\) and \(\bar M(T,\mu)\to\bar M(T)\)
    the suppression is systematically overestimated
    (the estimate for \(I_\text{CME}\) that we obtain is systematically lowered)
\begin{gather}
    I_\text{CME} \approx 8m^2\sum_{n=-\infty}^{+\infty}(-1)^{n+1}e^{-|n|\beta\frac{V_1}{2}+\mu n\beta} \times\\
    \times\int_0^{+\infty}s^2ds~\zeta(s)~
        \frac{e^{-\bar M^2s-\frac{(n\beta)^2}{4s}}}{(4\pi s)^2} \nonumber \\
    = \frac{m^2}{2\pi^2}\sum_{n=-\infty}^{+\infty}(-1)^{n+1}e^{-|n|\beta\frac{V_1}{2}+n\beta\mu} \times\\
    \times\frac{(n\beta)^2}{4}\int_0^{+\infty}\frac{dt}{t^2}~
        e^{-t-\frac{(\bar M n\beta)^2}{4t}} \nonumber 
\end{gather}
where we substituted the proper time \(s = \frac{(n\beta)^2}{4t}\) with a new dimensionless variable \(t\), and omit the regularization factor \(\zeta\).

\begin{figure}
    \center{\includegraphics[width=\linewidth]{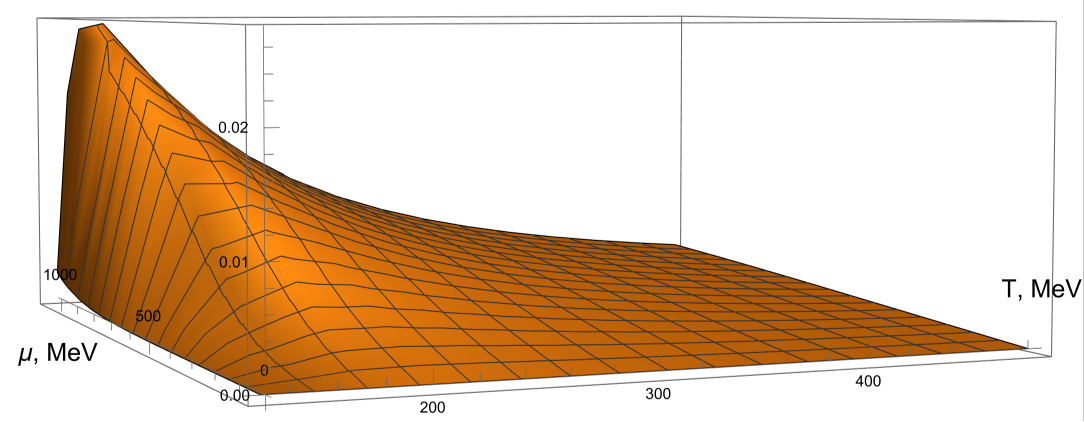}}
    \caption{
        \(I_\text{CME}(T,\mu)/I^\text{free}_\text{CME} = 2\pi^2I_\text{CME}(T,\mu)\) \eqref{EqCCME} for a light quark flavor, 
            where \(\mu=\mu_B/3\) is the quark chemical potential.
        The effect is the most pronounced in the phase diagram strip,
            where temperature \(T\) is just above the deconfinement transition,
            while the baryon density (chemical potential \(\mu\)) is as large as possible.
        For the non-perturbative inputs \(V_1(T),\bar M(T)\) values see the Appendix \ref{SectFCM}.
        The suppression is systematically overestimated at chemical potentials of order of nuclear and higher.
    }
    \label{FigICME1}
\end{figure}

With the Macdonald functions integral representations
\begin{align}
    K_\nu(z) &= \frac12(z/2)^\nu\int_0^{+\infty} e^{-t-\frac{z^2}{4t}}\frac{dt}{t^{\nu+1}}\\
    &= \frac{\sqrt\pi (z/2)^\nu}{\Gamma(\nu+\frac12)}\int_0^{+\infty} e^{-z\cosh t}(\sinh t)^{2\nu}dt 
\end{align}
\begin{gather}
    p = \bar M\sinh t, \quad \sqrt{p^2+\bar M^2} = \bar M\cosh t, \\
    dt = \frac{dp}{\sqrt{p^2+\bar M^2}} \nonumber \\
    \int_0^{+\infty}\frac{dt}{t^2}~e^{-t-\frac{(\bar M n\beta)^2}{4t}}
        = 4\int_0^{+\infty}\frac{dp~p^2 e^{-n\beta\sqrt{p^2+\bar M^2}}}{\bar M^2\sqrt{p^2+\bar M^2}} 
\end{gather}
we rewrite
\begin{gather}
    I_\text{CME} =
        \frac{1}{2\pi^2}\sum_{n=-\infty}^{+\infty}(-1)^{n+1}e^{-|n|\beta\frac{V_1}{2}+n\beta\mu} \times\\
    \times \frac{m^2}{\bar M^2}(n\beta)^2
        \int_0^{+\infty}\frac{dp~p^2 e^{-n\beta\sqrt{p^2+\bar M^2}}}{\sqrt{p^2+\bar M^2}}\nonumber 
\end{gather}
The factor \((n\beta)^2\) is produced by \(\frac{\partial^2}{\partial\mu^2}\),
    and the winding number sum adds up to the Fermi distribution \(f_\beta(E)=(1+e^{\beta E})^{-1}\)
\begin{gather}
    I_\text{CME}= \frac{1}{2\pi^2}\frac{m^2}{\bar M^2}
        \int_0^{+\infty}\frac{dp~p^2}{\sqrt{p^2+\bar M^2}}\Big(\nonumber \\
    f''_\beta\big(\sqrt{p^2+\bar M^2}+V_1/2-\mu\big) 
        + f''_\beta\big(\mu\to-\mu\big)\big) 
\end{gather}
and the final result is expressed via the integral emerging in CSE calculations \cite{VilenkinCSE,Metl,Zubkov2023}
\begin{gather}
    I_\text{CME} = \frac{-1}{2\pi^2}\frac{m^2}{\bar M^2(T)}\frac{\partial}{\partial\mu} \label{EqCCME}
        \int_0^{+\infty}\frac{dp~p^2}{\sqrt{p^2+\bar M^2(T)}}\Big( \\
    f'_\beta(\sqrt{p^2+\bar M^2(T)}+V_1(T)/2-\mu) 
    - f'_\beta(\mu\to-\mu)\Big) \nonumber
\end{gather}
For the values for the non-perturbative inputs \(V_1(T),\bar M(T)\) see the Appendix \ref{SectFCM}.
The result for a light quark flavor is plotted in Fig.\ref{FigICME1}.
Our analysis suggests that 
at high temperature CME in QCD is suppressed by the Color-Magnetic Confinement,
    parametrized with the function \(\bar M(T)\)
    that grows with temperature \eqref{EqMDnpQSE}.
At lower temperature the effect is suppressed by the ordinary Confinement, 
    remnants of which are parametrized with the function \(V_1(T)\) \eqref{EqPolPot}.

For a free light Dirac fermion the standard result is readily reproduced
    (for any temperature).
The integral was considered in \cite{VilenkinCSE,Metl}
    and is trivial at \(T\to 0\).
At \(V_1=0, ~|\mu| > m=\bar M\to 0\)
\begin{gather}
    I_\text{CME} = \frac{-1}{2\pi^2}\frac{m^2}{M^2}\frac{\partial}{\partial\mu}
        \int_0^{+\infty}\frac{dp~p^2}{\sqrt{p^2+m^2}}
        \Big(f'_\beta\big(\sqrt{p^2+M^2}-\mu\big) -\nonumber \\
    - f'_\beta\big(\sqrt{p^2+M^2}+\mu\big)\Big) 
    \xrightarrow[m=M\to 0]{} \\
    \frac{-1}{2\pi^2}\frac{\partial}{\partial\mu}
        \int_0^{+\infty}{dp~p}
        \Big(f'_\beta\big(p-\mu\big) - f'_\beta\big(p+\mu\big)\Big) 
        = \frac{1}{2\pi^2} \nonumber
\end{gather}
as expected.

\section{Conclusions and Discussions}

In this paper, we investigated the non-perturbative suppression of the Chiral Magnetic Effect.
With hot deconfined, yet still strongly interacting, QCD at non-zero baryon density 
    as a model for Quark-Gluon Plasma emerging in Heavy Ion Collisions, 
    we used the linear response to the chiral imbalance, 
    and retained only the linear order in external uniform magnetic field.
The Field Correlator Method was used to address the non-perturbative interaction in QCD. 
By quantifying the suppression of the CME through strong interaction effects, 
    we tried to advance the theoretical framework beyond perturbative approximations and phenomenological analysis, 
    offering a nuanced perspective on CME detectability across the QCD phase diagram. 

Our calculations revealed that the CME is significantly suppressed 
    across the most part of the QCD phase diagram, 
    challenging the expectation of robust anomalous transport in strongly interacting QCD matter. 
We found that the CME persists only in a narrow strip 
    (in the \(\mu_B-T\) phase diagram plane, see Fig.\ref{FigICME1})
    at baryon densities exceeding the nuclear density and temperatures just above the deconfinement transition. 
This behavior is in sharp contrast with our assessment of CVE and CSE \cite{Abramchuk2024CE},
    where the suppression is alleviated with either temperature and density.  

Beyond this strip, in particularly at higher temperatures, 
    the effect diminishes sharply due to the dominant influence of the Color-Magnetic Confinement, 
    which restricts the mobility of quarks essential for generating the chiral magnetic current.
However, our limited understanding of the gluonic dynamics at high baryon densities lead to quantitative overestimation of the suppression. 
With account of the effects of baryon density on the field correlators (on the singlet free energy \(V_1\) in particular), 
    we would expect the strip getting bent along the deconfinement transition line. 

These findings have significant implications for interpreting experimental signatures of the CME in heavy-ion collisions. 
The narrow window of persistence aligns with conditions probed by lower-energy collisions at 
    the upcoming FAIR, NICA and J-PARC-HI, 
    as well as at the existing SPS collider, 
    where high baryon densities and moderate temperatures supposedly prevail. 
In contrast, the suppression at higher temperatures, 
    typical for heavy ion collisions at LHC and RHIC
    suggests that detecting the CME in these regimes may be exceedingly difficult, 
    consistent with the inconclusive experimental results to date. 
However, in our analysis we did not discuss the characteristic magnitudes of the magnetic field and chiral imbalance 
    as functions of collision energy.
Nevertheless, our results thus refine the parameter space where anomalous transport might be observable, 
    addressing a challenge highlighted in the recent review \cite{Kharzeev2024CMEHIC}.

Upcoming CME analysis with Lattice QCD \cite{Brandt2025CMEKubo} might provide a validity test for our result, 
    though only in the zero baryon density limit.
Future work within FCM could explore the gluonic field correlators dynamic in baryon-dense medium.


\begin{appendices}
\begin{section}{Field Correlator Method for hot deconfined QCD}\label{SectFCM}

The Polyakov line in the deconfined region of the phase diagram
    incorporates the remnants of the Color-Electric interaction,
    which provides confinement in the confined region.
\(L\) is normalized in a way that its potential \(V_1\).
    \( L \approx \exp\left(-\frac{V_1(T)}{2T}\right) \)
is the energy required to overcome the remnant interaction 
    that bounds a heavy quark-antiquark color-singlet state.
The potential was not calculated within FCM so far, 
    so {we} rely on lattice data and use the fit \cite{Simonov2007jb}
\begin{gather}
    V_1(T>T_c) = \frac{175\text{ MeV}}{1.35 ~T/T_c - 1}, \label{EqPolPot}\\
    V_1(T_c) = 0.5\text{ GeV}, 
        \quad T_c = 160\text{ MeV}.\nonumber
\end{gather}
The potential depends on quark chemical potential via quark loops, 
    but {we} have to disregard the dependence, since it has not been investigated.

The spatial projection of the Wilson loop is defined by the Color Magnetic confinement,
CMC results in the Debye-like screening for quarks (and gluons) \cite{Agasian2006ra,Agasian2017,Andreichikov2017ncy}.
CMC grows with temperature, 
    which makes the Backgound Perturbation Theory for the thermal QCD self-consistent in higher orders of perturbation theory and resolves the Linde problem \cite{Simonov2016xaf}.
Meanwhile, the standard perturbation theory (the Hard Thermal Loop) provides no adequate suppression for high order diagrams.

The screened quark mass reads
\begin{gather}
    \bar M^2 = m^2 + \Big(\frac{c_D^2}{4}-\frac2\pi\Big)\sigma_{H,f}(T), \label{EqMDnpQSE}\\
    \quad\sigma_{H,f}(T)\approx c_\sigma^2g^4(T)T^2, \nonumber
\end{gather}
with the Debye-like screening in QGP \cite{Agasian2006ra} 
    (with \(c_D\approx 2\) \cite{Agasian2006ra} and \(c_\sigma\approx 0.56\) \cite{Simonov2022wcb}),
    and the non-perturbative spin-spin self-energy contribution \cite{Zubkov2023}.

The numerical constant entering the thermal running of the strong coupling, \(L_\sigma \approx 0.1\),
    in the one loop approximation \cite{Agasian2006ra}
\begin{align}
    g^{-2}(T) = 2b_0\log\frac{T}{T_cL_\sigma},
    (4\pi)^2b_0 = \frac{11}{3}N_c - \frac23 N_f.
    \label{EqGT}
\end{align}
The current quark mass running is numerically insignificant.

Assuming the Wilson loop in \eqref{EqPathInt} factorization \cite{Agasian2017} in spatial and temporal part,
the notion of singlet free energy allows us to estimate the temporal path integral (\(K_4 = \int_0^s d\tau \dot z_4^2/4\))
\begin{align*}
    \int({\cD z_4})_{0,n\beta}^s e^{-K_4} \times &\nonumber\\
    \times\langle N_c^{-1}\tr_c \Pexp &\left(ig\int B_4 dz_4 + \mu\int  dz_4\right)\rangle_B 
        \approx\nonumber\\
        &\approx \frac{e^{-\frac{n^2\beta^2}{4s}}}{\sqrt{4\pi s}}L^{|n|}\exp(\mu n\beta ).
\end{align*}
The notion of screened quark mass --- the spatial path integral
(\(K_3 = \int_0^s d\tau \dot{ \vec z}^2/4\))
\begin{gather*}
    \int({\cD^3\vec z})_{\vec x, \vec x}^s 
        e^{-K_3-m^2s} \exp(-\sigma_{H,f} S_3[\vec z(\tau)])
        \approx \frac{e^{-M^2s}}{(4\pi s)^{3/2}},
\end{gather*}
    and the non-perturbative spin-spin quark self-energy correction yields the negative shift \(\sim -\frac2\pi\) in \eqref{EqMDnpQSE} \cite{Zubkov2023}.

The field correlators and thus the string tension and the singlet potential essentially depend on baryon density.
The dependence is yet to be investigated.
Qualitatively, {we} expect the string tension and the singlet potential to decrease with chemical potential. 

\end{section}
\end{appendices}

\bibliographystyle{utphys} 
\bibliography{CMEinHotQGP.bib}

\end{document}